\documentclass[12pt]{article}
\usepackage{amssymb,amsmath}
\makeatletter

\@addtoreset{equation}{section}
\def\section{\@startsection {section}{1}{\z@}{-2.5ex plus -1ex minus
 -.2ex}{1.3ex plus .2ex}{\large\bf}}
\def\subsection{\@startsection{subsection}{2}{\z@}{-2.0ex plus%
 -1ex minus -.2ex}{0.3ex plus .2ex}{\bf}}

\advance \voffset by -0.6in
\advance \hoffset by -0.7in
\begin{document}
\parskip 7pt
\parindent 9pt

\centerline{\bf \Large  Michael Atiyah and Physics: the Later Years}

\begin{center}

{\bf  Bernd J~Schroers}\\
\vspace{0.2 cm}
Maxwell Institute for Mathematical Sciences and
Department of Mathematics,
\\Heriot-Watt University,
Edinburgh EH14 4AS, UK. \\
 {\tt b.j.schroers@hw.ac.uk}

\vspace{0.4cm}

{  August 2019} 
\end{center}

\begin{abstract}
\noindent I review aspects of Michael Atiyah's research in  fundamental theoretical physics during the last ten years of his life,  and contrast his deliberate interest in physics during this period with the  inadvertent impact on physics of his early mathematical work.  
\end{abstract}

\baselineskip 16pt
\parskip 4 pt
\parindent 10pt

\subsection*{The inadvertent and  the intentional physicist}

Michael Atiyah frequently expressed surprise at the extent to which his work on index theory and his early work in  gauge theory turned out to be important in physics. At the end of his commentary in  Volume 5 of his collected works he writes ``... I am really struck by the way most of the work which Singer and I did in the 60s and 70s has become relevant to physics".   Reading this work with hindsight one can indeed only marvel at  the contrast between the  absence of any physical motivation and the  ultimate importance  in  theoretical physics. It seems that, at least during the first half his mathematical career, Atiyah's impact on physics was entirely inadvertent. This is also borne out by a  remark in his 1998 lecture `The Dirac equation and geometry' where he  writes: ``When Singer and I were investigating these questions we `rediscovered' for ourselves the Dirac operator.  Had we been better educated in physics, or had there been the kind of dialogue with physicists which is now so common, we would have got there much sooner".

When Atiyah moved to Edinburgh in 1997,  his own physics education  had benefited hugely from  the dialogue between mathematicians and  physicists which he had done so much to initiate and develop. He was now interested in tackling fundamental questions in physics head-on, fully aware of their importance and with the actual intention of revolutionising the foundations of physics. The inadvertent physicist had become an intentional physicist.  

In this brief contribution I will collect some memories and impressions from Michael Atiyah's  second decade in Edinburgh, from  about 2009 until his  death in 2019. During this time we met regularly, sometimes weekly, for discussions related to the three main projects in physics which Atiyah pursued, namely his  conjecture regarding the configurations of points in three-dimensional Euclidean space,  the role of difference-differential equations in physics and geometric models of matter. 

Our  discussions had  started because the difference-differential equations which Atiyah had  introduced in the paper \cite{AM} with Greg Moore arise naturally in three-dimensional quantum gravity, which I was working on at the time. However, our focus soon shifted to  geometric models of matter  which became  the topic of   two joint papers \cite{AMS, AFS} and a jointly held EPSRC research grant\footnote{I believe this is Atiyah's only EPSRC  grant for personal research, i.e. not counting the ones he held as Director of the Isaac Newton Institute }.  I will focus on the geometric models here,  but more generally my goal is to identify  general  themes in Atiyah's thinking about physics. 

\subsection*{The inadequacy of conventional quantum mechanics}
Michael Atiyah expressed his dislike of conventional quantum mechanics on several occasions. He did not believe that any linear theory could be truly fundamental, and he shared Einstein's dislike of  the collapse of the wave function induced by observation. He felt that, in their efforts to go beyond quantum theory, physicists had tried many avenues but  had not critically questioned the  paradigm of initial value problems, i.e. the assumption that the future fields or wave functions can be computed from  values on a particular time slice. The difference-differential equation in \cite{AM} abandons  that assumption, and  introduces an element of non-locality. In his lectures on the subject, Atiyah highlighted these features as important motivations. 

\subsection*{The central role of the Dirac operator} 

The paper \cite{AM}  provides an elegant solution to the problem of  defining a finite time-shift operator  which is also relativistic by  exponentiating  the   Dirac operator. The use of the Dirac operator is no accident.     As we already saw,  Atiyah had `rediscovered' this operator in a purely mathematical context,  but in his later thinking about physics  it always played a central role, often as a possible bridge between the non-linear world of geometry and the linear world of quantum mechanics.  This was also  true in the work on geometric models of matter which I discuss next. 

\subsection*{The inadequacy of gauge theory and the  importance of  four-dimensional geometry} 

Atiyah's ideas for purely  geometric model of particles can be traced back to   the 1989 paper \cite{AtiyahManton}  with Nick Manton on the Skyrme model in nuclear physics. The paper is based on the observation that one can obtain three-dimensional  Skyrme fields by  computing the holonomy of  instantons in four dimensions, i.e. by integrating along the fourth dimension. Many years later,  work by Sakai and Sugimoto  \cite{SakaiSugimoto} and subsequent work by Sutcliffe \cite{Sutcliffe} suggested  that this holonomy can in fact  be viewed as the constant term in a Fourier series expansion of an instanton, and that  the other terms in this expansion  can  be attributed to further  three-dimensional  meson fields. The message was that three-dimensional physics could be described by a $SU(2)$ gauge theory in four (Euclidean) dimensions. 

Atiyah took this result  as an invitation to go one step further. He always felt that gauge theory, the mathematical language in which the standard model of elementary particles  is written, required too many arbitrary choices: of a  gauge group, of an associated vector bundle, of couplings and so on. His instinct was therefore to treat the appearance of a fourth dimension as an indication that  static, three-dimensional particles could be modelled in  terms of the geometry of four-manifolds. Formulating this in detail, and studying examples took several years and led to several publications, including \cite{AMS,AFS,AMa}. The basic challenge, only partly met, was to interpret the (integer) quantum numbers of particle  and nuclear physics - like baryon number, lepton number and electric charge - in terms of topological invariants of four manifolds. The Dirac operator featured, too, its kernel providing the linear space which encodes a particle's spin degrees of freedom.

There is no room here to discuss the geometric models of matter in any detail, but a few general observations about Atiyah's style of working during the collaboration on this topic may be of interest.  Even when he was well into his 80s, he worked hard and travelled frequently to attend conferences or to meet collaborators. When in Edinburgh he  came  to his office in the School of Mathematics at the University of Edinburgh almost every day and he put in additional shifts at home. He generated new ideas at a prodigious rate, announcing them with infectious enthusiasm but abandoning them casually if a new and more promising avenue presented  itself. These ideas were essentially mathematical  even though the questions  we discussed came from physics. Atiyah needed discussion partners, or at least listeners,  to develop his ideas, and he was usually juggling several projects and associated conversations at any point in time. Remarkably, he always took a personal interest in his discussion partners  and created additional connections where he could.  For example, I was  travelling to Africa   regularly during our collaboration to teach at one of the African Institutes for Mathematical Sciences (AIMS). Atiyah immediately  offered his help, and joined the international advisory board for the AIMS centre in Ghana.

\subsection*{Division algebras and Bohmian mechanics}

In our discussions, Atiyah frequently mentioned two other themes which he wanted to incorporate in his description of matter, namely division algebras and Bohmian mechanics.  They never entered his published work on physics in any detail, but I think they are worth recording.  Atiyah felt that the  four division algebras - real and complex numbers, quaternions and octonions  - provided essentially the only  mathematically natural way to  account for  the number of fundamental forces (four) or the number of generations (three) in the standard model. However, he never settled on a definite proposal of how this matching could work.

Atiyah  liked David Bohm's formulation of quantum mechanics in terms of non-local classical variables which follow trajectories determined by the wave function. He felt it clarified  some of the foundational issues in quantum mechanics by making the non-locality explicit,  and he  hankered after a role for them in his ideas about particles. We  discussed a possible use related to his configuration space conjecture, but never came up with a convincing proposal.

\subsection*{Passion and beauty}

Anybody who interacted with Michael  Atiyah soon noticed that he had a strong personal taste in life and in science, and that his instincts were not easily discouraged. Sometimes I felt Atiyah did not take on board the painstaking work required to match a theoretical model in physics to the experimental data, and  occasionally our   discussions would  turn into arguments. On one such occasion, having listened to Atiyah's latest ideas,  I asked ``But this is just a gut feeling, right?",  to which he shot back ``Yes, but it is {\em my} gut!".   Presumably his awareness that his early work, motivated by  strictly mathematical considerations, had been so unexpectedly and powerfully relevant to physics boosted his confidence, but I should stress that I never discussed this with him.

Michael Atiyah trusted his own instincts, but  he  also trusted great minds. He had his heroes, and in physics these were, above all,  Maxwell,  Einstein and Dirac. He thought that all their ideas would ultimately prove right in some sense.  With Einstein and Dirac he was aware that the work they had done as young men was celebrated whereas aspects of their  later work (e.g. Einstein on unified field theory  and Dirac on large numbers) were viewed critically. He felt this was not justified, and saw parallels in his own biography.

In an interview for  the Newsletter of the European Mathematical Society in November 2018, the last one he gave, Atiyah offered the following advice  to young mathematicians: ``What you need is passion, persistence and risking things for searching for the beauty". He certainly lived that advice. 
Atiyah's courage to address fundamental questions in science head-on, to follow his own instincts and dreams in seeking answers, and to use beauty as a guide  meant that discussions with him  transcended the humdrum of scientific activity. Atiyah sought and encouraged a similar ambition and independence of mind in others, too. In doing so, he took a risk and sometimes attracted criticism, particularly in later years. However, he also created a space for both deep and lateral thinking which is rare  and which many  treasured. 

\vspace{0.5cm}

\noindent {\bf Acknowledgements} \; I thank Nigel Hitchin for inviting me to write up my recollections of interacting with Michael Atiyah and  Jos\'e Figueroa O'Farrill for comments on an  earlier version of this article.

\end{document}